\documentclass{article} 
\usepackage{url} 
\usepackage{amsthm}
\usepackage{amsmath}
\usepackage{amssymb}

\usepackage{tikz}
\usetikzlibrary{calc}
\usetikzlibrary{matrix}
\usetikzlibrary{positioning}

\title{A Game Theoretic Analysis of Liquidity Events in Convertible Instruments} 

\author{R. van der Meyden\\ 
School of Computer Science and Engineering,\\
 UNSW Sydney\\ 
R.VanderMeyden@unsw.edu.au} 

\newtheorem{thm}{Theorem}

\theoremstyle{definition}
\newtheorem{example}{Example}

\newcommand{\Safes}{\mathbf{C}} 
\newcommand{\Converted}{C} 
\newcommand{\principal}{p} 

\newcommand{\Cash}{K} 

\renewcommand{\Cap}{c} 
\newcommand{\shares}{s} 

\newcommand{\scommon}{\shares_{\mathit{common}}}
\newcommand{\sconverted}[1]{\shares(#1)} 
\newcommand{\liquidityCap}[1]{S(#1)}
\newcommand{\liquidityPrice}[1]{P(#1)}  
\newcommand{\Value}{V} 
\newcommand{\CashoutValue}[1]{\mathtt{cash}(#1)}
\newcommand{\ConversionValue}[1]{\mathtt{conv}(#1)}
\newcommand{\payout}{U}
\newcommand{\sfound}{s}

\newcommand{\Reals}{\mathbb{R}}
\newcommand{\pow}[1]{{\mathcal P}(#1)}

\newcommand{\nashord}{\preceq} 

\newcommand{\commentout}[1]{}

\begin{document} 
\maketitle 

\begin{abstract} 
Convertible instruments are contracts, used in venture financing, which give investors 
the right to receive shares in the venture in certain circumstances. In 
liquidity events, investors may have the option to
either receive back their principal investment, or to receive a proportional payment 
after conversion of the contract to a shareholding. In each case, the value of the payment 
may depend on the choices made by other investors who hold such convertible contracts. 
A liquidity event therefore sets up a game theoretic optimization problem. The paper defines 
a general model for such games, which is shown to cover all instances of the Y Combinator 
Simple Agreement for Future Equity (SAFE)
contracts, a type of convertible instrument that is commonly used to finance startup ventures. 
The paper shows that, in general, 
pure strategy Nash equilibria do not necessarily exist in this model, and there may not exist an optimum pure strategy Nash 
equilibrium in cases where  pure strategy Nash equilibria do exist. 
However, it is shown when all contracts are uniformly one of the SAFE contract types, an optimum pure strategy Nash equilibrium exists. 
Polynomial time algorithms for computing (optimum) pure strategy Nash equilibria in these cases are developed. 
\end{abstract} 


\section{Introduction} 

Convertible instruments are a form of contract between an investor 
and a company, that provide the holder an option to 
convert the contract to shares in the company in certain circumstances. 
Examples include convertible bonds, which provide an interest stream in addition 
to the conversion option. Simpler instruments such as Y Combinator's Simple Agreements for 
Future Equity (SAFEs) \cite{YC-old,YC} are common in seed financing of startups, and provide only 
the conversion rights. Our focus in this paper is on this simpler form of contract.  

Startup shares generally do not trade publicly.  ``Liquidity Events'', as defined in SAFEs, 
are situations where the company undergoes a change of control, direct listing, or initial public offering, 
providing early investors an exit opportunity. 
In a Liquidity Event, the SAFE holders
have an option to receive either cash (the Cashout Amount), 
or to convert the SAFE to a certain number of shares, the value 
of which is called the Conversion Amount. 
(In acquisitions, the proceeds to be distributed to the investors 
may be money and/or acquirer shares, but we focus in this paper on the corresponding monetary value.) 

The Liquidity Event clause in some SAFE contracts states that the 
SAFE holder gets the maximum of the Cashout Amount and the Conversion amount. 
However, these amounts may depend on the choices made by other investors, so it is unclear 
that there is a determinate maximum. This is so particularly as the Cashout rights are 
senior to the Conversion rights, so that moneys to be 
paid out to the holders of shares received in conversion first have the Cashout amounts 
deducted. 

SAFE contracts therefore create a game theoretic scenario 
in which the players (of some finite number) are the holders of SAFE contracts, and the moves are 
to choose to Cashout or to Convert. The details of the payouts depend on the 
specific type of SAFE, of which there are multiple versions, depending on a number of 
parameters. SAFEs may or may not include a ``Cap'', a maximum valuation of the 
company at which the conversion will be calculated. Additionally, there may or may not be a ``Discount"  
on the market price of the shares applied in the event of a conversion. Finally, the valuation used in the 
conversion calculation may be a ``Pre-Money Valuation" or a ``Post-Money Valuation". 
(Y Combinator's orginal SAFES used the Pre-Money Valuation method, but they have recently switched 
to the Post-Money method.). 

Our contributions in this paper are the following. 
\begin{itemize} 
\item  In Section~\ref{sec:game-model}, we define 
a general game model for liquidity events in convertible instruments. 

\item We show in Section~\ref{sec:safes}  that 
the general game model covers all SAFE variants, in the sense that for each variant, if 
all SAFE contracts are of the same type, then the general model captures the resulting game in 
Liquidity Events. 
\item In Section~\ref{sec:general} we show that, in 
general, the game is not guaranteed to have a pure strategy Nash equilibrium. Moreover, 
where such equilibria exist, they may be in conflict, with none maximizing the payout to all 
players.  However, we also identify a sufficient condition that ensures that, if a pure strategy Nash equilibrium exists, there exists
an optimum such equilibrium, that maximizes the payout to all players. All the SAFE contract variants satisfy this sufficient 
condition. 

We develop a polynomial-time algorithm 
to compute a set of representatives of the equivalence classes of 
the pure strategy Nash equilibria and the optimum such equilibrium. 
(The equivalence relation makes two equilibria equivalent if for each player, the payout is the same.)
Note that the number of game matrix entries for an $n$-player game in normal form 
is exponential, but we describe the input game using just a constant set of numerical parameters for each 
player - the algorithm is polynomial in the total length of such inputs. 

\item In Section~\ref{sec:nash-pre-money} we show that the Pre-Money SAFE variants 
in fact have a pure strategy Nash equilibrium. We show that in this case, an optimization of the algorithm for computing 
representative equilibria is possible. We also show by example that there may be an exponential number of equivalent 
optimum equilibria. 

\item  In Section~\ref{sec:nash-post-money} we show that, similarly,  the Post-Money SAFE variants 
also in fact have a pure strategy Nash equilibrium. A different proof is
required for this case, however. The algorithm for computing equilibria, in this case, computes not just a
representative set of equilibria, but all equilibria. 

\item In Section~\ref{sec:mixed} we show by an example that pure strategy Nash equilibria of the Liquidity Event game are not guaranteed to exist when the company has issued both Pre-Money SAFE contracts and Post-Money SAFE contracts. 

\end{itemize} 
The fact that all SAFE variants (provided they are used uniformly and not mixed) result in the Liquidity Event game having 
an optimum pure strategy Nash equilibrium shows that the reference
to the maximum of the Cashout Amount and Conversion amount in some SAFE variants 
is sound, since this maximum can be interpreted as the payout in received by the SAFE holder in 
the optimum equilibrium. 
Moreover, it shows that the Liquidity Event clauses in these contracts are well-designed, in the sense that they 
do not create inherent conflicts between the investors, that would need to be resolved 
by extra-contractual means. 

Section~\ref{sec:concl} discusses related work and open problems arising from these results. 

\section{The Liquidity Event Game} \label{sec:game-model}

We consider a liquidity event, carried out under the following assumptions. 
An amount of value $\Value>0$ is to be paid out to the investors in the company. We assume that all 
debts of the company and claims senior to the shareholders and holders of convertible instruments 
have already been extinguished, so that $\Value$ represents the amount of value to be distributed to the 
investors with claims on equity in the company. We assume that $\Value$ consists of monetary value, but 
the analysis would be essentially the same if it were denominated in some other form that can be valued in monetary terms (for example, shares of the acquirer in case of an acquisition of the company.)  

We suppose that a nonempty finite set $\Safes$ of convertible instruments has been issued by the company, 
with instrument $i\in \Safes$ purchased for amount $\principal_i >0$, the ``principal''. In a liquidity event, 
each instrument gives the investor a choice between ``cashing out" or ``converting". 
In case of a cashout, the investor recovers their principal (or a pro-rata amount, in case there is 
insufficient value to be distributed). In case of a conversion, the instrument is first converted to 
shares, and the value remaining is distributed in proportion to the shareholdings after conversion.  
Cashouts are senior to conversions, so that the principal amounts are repaid to investors  choosing to Cashout,
before a distribution is made based on existing shareholdings and shares issued in conversion. 
We assume that there are no preferred shareholders, and that common stockholders 
are treated at the same priority as converting investors.%
\footnote{The assumption that there are  no preferred shares is reasonable in the case of SAFEs since
these instruments are generally used prior to the first priced equity round, and are converted in that round.}

Thus, the set of players of the game is $\Safes$, and each $i\in \Safes$ has two moves Cashout and Convert.
We may therefore represent a strategy profile $\sigma: \Safes \rightarrow \{{\rm Cashout}, {\rm Convert}\}$
by the set  $\Cash \subseteq \Safes$  of instruments that are cashed out, 
that is, the set of $i\in \Safes$ such that $\sigma(i) = {\rm Cashout}$. 
Conversely, 
$\overline{\Cash} = \Safes \setminus \Cash$ is the set of instruments that convert. 
We describe the payouts $\payout_i(\Cash)$ for each player $i \in \Safes$ when 
$\Cash$ is the set of instruments that are cashed out. 

For $i \in \Cash$, the payouts are straightforwardly defined from the principal amounts. 
Define
$$\principal(\Cash) = \sum_{i\in \Cash} p_i$$
to be the total amount of principal paid in for the instruments that are being cashed out. 
The payouts for $i \in \Cash$ are then defined by 
$$ \payout_i(\Cash) = \left \{ \begin{array}{lr}  
\principal_i  & \mbox{if $\principal(\Cash) \leq \Value$} \\  
\frac{\principal_i}{\principal(\Cash)} \cdot \Value & \mbox{otherwise}
\end{array} \right.
$$   
That is, investors cashing out receive their principal amount, if the funds $V$ to be distributed are enough to cover
such payments, or a pro-rata share of $V$ otherwise.

We write $\CashoutValue{\Cash}$ for $\sum_{i\in \Cash} \payout_i(\Cash)$, the total payout to 
investors that take the cashout option. 
This means that the amount to be distributed to shareholders after 
the conversions is $V- \CashoutValue{\Cash}$.  This amount is greater than or equal to zero, and equal to zero if 
$\principal(\Cash) \geq V$. 

We assume that the payouts to investors $i \in \overline{\Cash}$ taking the conversion option is defined by
$$\payout_i(\Cash) = \frac{\principal_i}{\gamma_i f(\overline{\Cash})} \cdot (V- \CashoutValue{\Cash})$$
where 
\begin{itemize} 
\item $f: \pow{\Safes} \rightarrow \Reals$ is a function that is linear in the sense that for $X\subseteq \Safes$, we have 
$$f(X) = \beta+ \sum_{i\in X} \alpha_i$$
where $\beta$ and the $\alpha_i$ are positive constants, and 
\item $\gamma_i$ is a positive constant.
\end{itemize} 
Thus, the payout to investor $i$ is proportional to the principal $\principal_i$ paid in by investor $i$, 
divided by a number $f(\overline{\Cash})$ that depends on the set $\overline{\Cash}$ 
of investors converting, and a 
constant factor $\gamma_i$ that depends only on the investor $i$. 
Plainly, for this distribution to be well-defined, 
we require that for all non-empty 
$\Converted = \overline{\Cash} \subseteq \Safes$, with we have 
\begin{equation} 
\sum_{i\in \Converted} \frac{\principal_i}{\gamma_i f(\Converted)}  < 1 ~. 
\label{eq:game-well-defined}
\end{equation} 
Strictness of the inequality comes from 
taking into account that the company will have a non-zero number of shares before adding converted shares.
The restriction to non-empty set $\Converted$ is because in the case $\Converted =\emptyset$, 
there is no distribution to be made to holders of convertible instruments. 

In particular, taking $\Converted = \{i\}$, we
see that 
\begin{equation} 
\principal_i < \gamma_i f(\{i\}) = \gamma_i(\beta + \alpha_i)\label{eq:game_well-defined-i}  
\end{equation} 
is a necessary condition for the distribution to be well defined. 

A further constraint for technical reasons to be explained below is the following inequality: for each $i\in \Safes$, 
\begin{equation} 
\gamma_i \alpha_i \leq \principal_i \label{eq:solvable}
\end{equation} 

We will show below that the above model is sufficiently general to cover the liquidity event provisions of all versions of the Pre-Money and Post-Money SAFE contracts.

\section{SAFE Contract Variants} \label{sec:safes} 

SAFE contracts come in multiple forms. As originally issued \cite{YC-old}, the conversion definitions in the event of an equity round 
were based on the new investor's pre-money valuation of the company. More recent versions \cite{YC}  are based on the post-money valuation. Both Pre-Money and Post-Money SAFEs come in four forms, depending on whether a 
cap and/or a discount are included as parameters of the contract.%
\footnote{This gives four variants of each of the Pre-Money and  Post-Money version, but the 
``Cap and Discount" version of the Post-Money SAFE is no longer provided at  \cite{YC}, so we describe only the 
three versions of the Post-Money SAFE that have been provided.
(The reason for the omission of the Post-Money SAFE with Cap and Discount appears to be that a conversion formula 
for this SAFE in Liquidity Events was not included in the contract previously published.)
}

The Post-Money SAFE variants specify that the investor receives the maximum of the Cashout Amount and the Conversion Amount.  As already noted, given the game theoretic structure of the situation, the 
interpretation and existence of a maximum need to be justified, so we describe the payouts for each of the 
options and seek to prove the existence of an optimum that can be used to interpret this maximization 
requirement. 

In both cases, the variant with neither a cap nor a discount (also known as a Most-Favoured Nation (MFN) SAFE due to included protections in case more favourable terms are granted to a later investor) does not raise the game theoretic issues that are our focus in this paper. This variant simply states that in a liquidity event, the investor receives their principal back (or a pro-rata share in case of insufficient funds). That is, the only option for the investor is to cash out, as described above. 
In the event that the company issues another SAFE variant (after issuance of the MFN SAFE, but before termination 
of the contract), the investor has an option to convert their contract to the terms of the later issued SAFE. 
The analyses of these other versions can therefore be applied in the event that this option is taken up. 

We show in this section that the liquidity event provisions of the remaining SAFE variants of 
both types are instances of the general game model of Section~\ref{sec:game-model}. In all these variants, 
the distribution in the case of investors choosing to cash out is as described above for the case $i \in \Cash$, so we focus on the case where $i \in \overline{\Cash}$. 

We assume as a simplification that there are no preferred shares, options or other types of convertible instruments, so that the distribution is to  SAFE holders and common stock holders only. We suppose also that all SAFEs issued by the company are of the same type, though they may differ in their parameter settings.  
The Liquidity Event clauses of these contracts have qualifications relating to the 
form of the distribution (e.g., a mix of cash and shares), and provisions relating to tax-free reorganisations, 
which we also omit for simplicity.

\subsection{Pre-Money SAFE with Cap Only} \label{sec:safe-cap}

The Pre-Money SAFE with Cap Only has as parameters the principal (purchase amount) $\principal_i$
and a Valuation Cap $\Cap_i$. In a liquidity event, this SAFE offers the investor the option to cash out or to convert. 
The number of shares issued in conversion of the SAFE 
is defined as the principal amount $\principal_i$ divided by the Liquidity Price. 
Liquidity Price is defined as ``the price per share equal to the Valuation Cap divided by the Liquidity Capitalization".
The Liquidity Capitalization, in turn is defined by 
\begin{quote}
``{\bf Liquidity Capitalization}'' means
the number, as of immediately prior to the Liquidity Event, of shares of Capital Stock (on an as-converted basis) outstanding, assuming exercise or conversion of all outstanding vested and unvested options, warrants and other convertible securities, but excluding: (i) shares of Common Stock reserved and available for future grant under any equity incentive or similar plan; (ii) this instrument; (iii) other Safes; and (iv) convertible promissory notes.
\end{quote} 
We write $\sfound$ for the liquidity capitalization. Thus, we 
obtain that the Liquidity Price for SAFE $i$ is equal to $\Cap_i/\sfound$, and the number of shares issued in conversion
is equal to $\principal_i \sfound/\Cap_i$. After conversion, the total number of shares in the company is therefore 
$ \sfound + \sum_{i\in \overline{\Cash}} ~\principal_i \sfound/\Cap_i$, and we obtain that the share of investor $i \in \overline{K}$ of the 
distribution is 
\begin{align*}
\payout_i(\Cash) & = \frac{\principal_i \sfound/\Cap_i}{ \sfound + \sum_{j\in \overline{\Cash}} ~\principal_j \sfound/\Cap_j} \cdot (V - \principal(\Cash)) \\ 
 & = \frac{\principal_i}{ \gamma_i f(\overline{\Cash})} \cdot (V - \principal(\Cash))
 \end{align*}
where $\gamma_i = \Cap_i$ and $f(X) = 1 +\sum_{j\in X} \principal_j /\Cap_j$. 
This distribution therefore fits the Liquidity game model with $\beta = 1$ and $\alpha_i = \principal_i/\Cap_i$.

It is immediate from the definitions that $\sum_{i\in \overline{K}} \payout_i(K) < V - \principal(K)$, 
so condition~(\ref{eq:game-well-defined}) holds. 
Note also that $\gamma_i \alpha_i = \principal_i$ in this case, so inequality~(\ref{eq:solvable}) is satisfied. 

\subsection{Pre-money SAFE with Discount Only} \label{sec:discount}

\newcommand{\discount}{d} 
\newcommand{\price}{\pi}  

A Pre-Money SAFE with Discount Only has as parameters a principal amount $\principal_i$ and a Discount Rate 
$\discount_i\leq 1$. Using this, the Liquidity Price is defined by 
\begin{quote}
``{\bf Liquidity Price}'' means the price per share equal to: the fair market value of the Common Stock at the time of the Liquidity Event, as determined by reference to the purchase price payable in connection with such Liquidity Event, multiplied by the Discount Rate. 
\end{quote} 
Thus, the Liquidity Price is for SAFE $i$ is $\price \discount_i$ where $\price$ is provided as an input to the liquidity event. 
The number of shares issued in conversion is again the  principal divided by the Liquidity Price, giving  
a  share of the distribution for SAFE investor $i$ of 
\begin{align*}
\payout_i(\Cash) & = \frac{\principal_i /\price\discount_i}{ \sfound + \sum_{j\in \overline{\Cash}} ~\principal_j /\price\discount_j }\cdot (V - \principal(\Cash)) \\ 
 & = \frac{\principal_i}{ \gamma_i f(\overline{\Cash})} \cdot (V - \principal(\Cash)
 \end{align*}
where $\gamma_i = \price\discount_i$ and $f(X) = s +\sum_{j\in X} \principal_j /\price\discount_j$. Here we have $\beta = s$ and $\alpha_j = \principal_j /\price\discount_j$. Hence, again, 
$\gamma_i \alpha_i = \principal_i$ and  inequality~(\ref{eq:solvable}) is satisfied. 
It is immediate from the definitions that 
condition~(\ref{eq:game-well-defined}) holds.

\subsection{Pre-Money SAFE with Cap and Discount}  \label{sec:cap-discount}

A Pre-Money SAFE with Cap and Discount has as parameters the principal $\principal_i$, 
a cap $\Cap_i$ and a discount rate $\discount_i$. In the event of an equity round, these are used to select the 
most favourable conversion outcome from the conditions from the Pre-Money SAFE with Cap Only and the Pre-Money
SAFE with Discount Only. 
However, in a Liquidity Event, the conversion conditions from the Pre-Money SAFE with Cap Only are applied. The definitions for this contract are therefore similar to those given in Section~\ref{sec:safe-cap}.

\subsection{Post-Money SAFE with Cap Only} \label{sec:post-money-cap}

The Post-Money SAFE with Cap Only is written in a way that does not directly describe a conversion formula, but instead sets up a set of simultaneous equations that need to be solved in order to derive the number of shares into which the contract converts. 
In effect, the Post-Money SAFE grants the SAFE investor a 
fixed proportion of the company in certain circumstances. This is the case even when other SAFE investors are
granted shares in the company in conversion of their SAFEs, which might otherwise dilute the SAFE holder - the interaction of these effects leads to the equations that need to be solved. 
This is explained elsewhere for the Equity Financing clause \cite{MeyMa20}. A similar issue arises for 
the Liquidity Event clause in this contract. 

We explain the derivation here for the Post-Money SAFE with Cap Only. 
The parameters of this contract, like the Pre-Money version, are
the principal  $\principal_i$ paid for SAFE $i$, and a cap 
$\Cap_i$ that is called the Post-Money Cap of the SAFE.
 
For a Post-Money SAFE to be meaningful and well-defined, we need $\principal_i < \Cap_i$, 
because on conversion, the Post-Money SAFE is converted to a number of shares that gives the 
SAFE investor a $\principal_i/\Cap_i$ share of the company (in the case of a conversion triggered by an equity round, this share is prior to the new issuance for the new money). 
For the same reason, we need that the total conversion share of all SAFES is less than 1, so that the founders 
are left with a non-zero share of the company. That is, we need $\sum_{i\in \Safes}~ \principal_i/\Cap_i < 1$. 

The Liquidity Event clause of the SAFE says
\begin{quote} 
{\bf Liquidity Event.}  If there is a Liquidity Event before the termination of this Safe, this Safe will automatically be entitled to receive a portion of Proceeds, due and payable to the Investor immediately prior to, or concurrent with, the consummation of such Liquidity Event, equal to the greater of (i) the Purchase Amount (the ``Cash-Out Amount'') or (ii) the amount payable on the number of shares of Common Stock equal to the Purchase Amount divided by the Liquidity Price (the ``Conversion Amount'').  
\end{quote} 
Compared to the Pre-Money version, we might note that this assumes that the greater of (i) and (ii) is well-defined. 
Because of the dependence of these values on other SAFEs, this is in fact not clear, so we will model 
this maximization instead as a choice of the investor, and derive the existence of an optimum choice. 

The Liquidity Price and Liquidity Capitalization are defined by 
\begin{quote} 
``{\bf Liquidity Price}'' means the price per share equal to the Post-Money Valuation Cap divided by the Liquidity Capitalization.\\~\\
``{\bf Liquidity Capitalization}'' is calculated as of immediately prior to the Liquidity Event, and (without double- counting): 
\begin{itemize} 
\item	Includes all shares of Capital Stock issued and outstanding;
\item Includes all (i) issued and outstanding Options and (ii) to the extent receiving Proceeds, Promised Options;
\item Includes all Converting Securities, other than any Safes and other convertible securities (including without limitation shares of Preferred Stock) where the holders of such securities are receiving Cash-Out Amounts or similar liquidation preference payments in lieu of Conversion Amounts or similar ``as-converted'' payments; and
\item Excludes the Unissued Option Pool.
\end{itemize} 
\end{quote} 
SAFEs are a form of Converting Security. 
Thus, whereas the Liquidity Capitalization in the case of the Pre-Money SAFEs was a constant, this
definition is a function of the set of SAFEs not already cashed out. 

To formalize these definitions, suppose $\Converted = \overline{\Cash}$ 
is the set of SAFES that are being converted, and let $i \in \Converted$. 
Relative to these parameters, let 
\begin{itemize} 
\item $\sconverted{\Converted}_i$ be the number of shares issued to SAFE investor $i$ in conversion of their SAFE,  
\item $\liquidityCap{\Converted}$ be the Liquidity Capitalization,
\item $\CashoutValue{\Converted}_i$ be the Cashout Amount of the SAFE $i$, 
\item $\ConversionValue{\Converted}_i$ be the Conversion Amount of the SAFE $i$, and 
\item $\liquidityPrice{\Converted}_i$ be the Liquidity Price of SAFE $i$. 
\end{itemize} 
Note that these values may depend on their parameters. 

The conversion value is determined as follows. We assume that there are no convertible securities other than SAFEs, 
so write $\scommon$ for the (constant) number of common shares after all outstanding options and warrants have 
been vested. Then 
$$\liquidityCap{\Converted} = \scommon + \sum_{i\in \Converted} \sconverted{\Converted}_i$$
by the definition of Liquidity Capitalization. However, $\liquidityCap{\Converted}_i$ in turn depends on $\sconverted{\Converted}$, so we have a circularity.
Specifically, we have that, the Liquidity Price is defined as 
$$\liquidityPrice{\Converted}_i = \Cap_i/ \liquidityCap{\Converted} $$ 
and, when $i \in \Converted$, that $ \sconverted{\Converted}_i $ is defined as the purchase amount divided by the Liquidity Price, 
i.e. 
$$ \sconverted{\Converted}_i = \principal_i/ \liquidityPrice{\Converted}_i$$ 
Combining these equations, we get 
\begin{equation} 
\sconverted{\Converted}_i = (\principal_i/ \Cap_i) \liquidityCap{\Converted}
\label{eq:sci}
 \end{equation} 
 and hence 
 $$\sconverted{\Converted}_i = (\principal_i/ \Cap_i) (\scommon + \sum_{i\in \Converted} \sconverted{\Converted}_i)
$$
 which displays the circularity via $ \sconverted{\Converted}_i$ (we could also display it via $ \sconverted{\Converted}$).
 
However, we do not actually need to compute $\sconverted{\Converted}_i$ in order to determine the Cashout and Conversion amounts for a given choice of $\Converted$. 
Clause (d), ``Liquidation Priority" states that the 
conversion amounts are junior to cashout amounts, that is, cashouts are paid first, and
the remaining value is issued to common and converted shares. 
For $i \in \Cash$, the Cashout Amounts are identical to the amounts $\payout_i(\Cash)$ of Section~\ref{sec:game-model}. 

The Conversion Value of SAFE $i$ is the share of the total Conversion amount corresponding to the 
proportional shareholding after conversion, that is, 
$$
\begin{array}{rlr}  
\ConversionValue{\Converted}_i  & = \frac{\sconverted{\Converted}_i}{\sconverted{\Converted}}
\cdot (V-\CashoutValue{\Cash}) \\[10pt] 
                                                    & = \frac{\principal_i}{\Cap_i}\cdot   (V-\CashoutValue{\Cash})&
                                                      \end{array}
$$ 
by equation~(\ref{eq:sci}). 
The payout to investor $i \in C=\overline{K}$ is $\payout_i(K) = \ConversionValue{\Converted}_i  $. 
This fits the game model of Section~\ref{sec:game-model} with 
$\gamma_i = \Cap_i$ and $f(X) = 1$, that is, $\beta= 1 $ and $\alpha_j = 0$ for all $j \in \Safes$. 
Note also that $\gamma_i \alpha_i - \principal_i = -\principal_i <0$, so inequality~(\ref{eq:solvable}) is satisfied.  
We have already stated that $ \sum_{j\in \Safes}~ \principal_i/\Cap_i <1$ as an assumption: it follows from this that 
condition~(\ref{eq:game-well-defined}) holds.

The above derivation does leave open the question of whether $\sconverted{\Converted}_i$ has been properly defined, given the circularity. 
This can be addressed by noting that the circularity is easily resolved by treating these definitions as simultaneous equations. Summing equation~(\ref{eq:sci}) over $i \in \Converted$, we get 
$$ \sum_{i \in \Converted} \sconverted{\Converted}_i  =   \left(\sum_{i \in \Converted} \principal_i/ \Cap_i\right) \left(\scommon + \sum_{i\in \Converted} \sconverted{\Converted}_i\right)$$ 
which we can reorganise to get a closed form solution for the total converted shares: 
$$ \sum_{i \in \Converted} \sconverted{\Converted}_i  =   \frac{\sum_{i \in \Converted} \principal_i/ \Cap_i }{ 1 - \sum_{i\in \Converted} \principal_i/\Cap_i} \cdot \scommon$$ 
Substituting this back into the above definition of the Liquidity Cap, we get 
$$
\begin{array}{rl}  
\liquidityCap{\Converted}  & = \scommon +    \frac{ \sum_{i\in \Converted}  \principal_i/ \Cap_i}{ 1 - \sum_{j\in \Converted} \principal_j/ \Cap_j}  \cdot \scommon  \\[10pt] 
& =  \frac{\scommon}{ 1 - \sum_{j\in \Converted} \principal_j/ \Cap_j}  
\end{array} 
$$
Also, substituting into equation~(\ref{eq:sci}), we get 
$$ \begin{array}{rl}
\sconverted{\Converted}_i & =  (\principal_i/ \Cap_i) \liquidityCap{\Converted} \\[10pt] 
& = \frac{\principal_i/ \Cap_i}{ 1 - \sum_{j\in \Converted} \principal_j/ \Cap_j}  \cdot \scommon 
\end{array} 
$$ 
which shows that the notion of Conversion Payout is well-defined, given $\Converted$.

\subsection{Post-Money SAFE, Discount Only} \label{sec:postdiscount}

The Post-Money SAFE with Discount Only, like the Pre-Money version, has a discount $d_i$ as a parameter of the 
contract. The payouts in a Liquidity event are exactly as described for the Pre-Money SAFE with Discount Only 
in Section~\ref{sec:discount}. Hence, equation 
$\gamma_i \alpha_i = \principal_i$ and  inequality~(\ref{eq:solvable}) are satisfied for this SAFE variant.

\section{Nash Equilibria in the General Game} \label{sec:general}

The SAFE contracts' handling of Liquidity Events makes the most sense if it can be proved that there is always a choice of moves that is the best possible for all players.

Given a Liquidity Event game on a set of players $\Safes$, which determines payouts $\payout_i(\Cash)$ for $i \in \Safes$, 
define the binary relation $\nashord$ on the set of strategy profiles $\pow{\Safes}$ by $\Cash_1 \nashord \Cash_2$ if 
for all $i \in \Safes$ we have $\payout_i(\Cash_1) \leq \payout_i(\Cash')$. 
It can be seen that this relation is a quasi-order, i.e., it is reflexive and transitive. 
We write $\Cash \approx \Cash'$ when both $\Cash \nashord \Cash'$ and $\Cash' \nashord \Cash$. 
Intuitively, $\Cash \approx \Cash'$ when $\Cash$ and $\Cash'$ yield the same payouts for all players. 
(We will see below that this is possible even when $\Cash \neq \Cash'$.)

An \emph{optimum} strategy profile 
is a strategy profile $\Cash^*$ such that $\Cash \nashord \Cash^*$ for all strategy profiles $\Cash$. 
Optima, when they exist, are not necessarily unique, since we may have two distinct optima $\Cash \approx \Cash'$.

In general, it is not clear that an optimum exists. A weaker
notion is the following.  A strategy profile $\Cash$ is a 
\emph{pure strategy Nash equilibrium}
 if for all players $i$, we have $\payout_i(\Cash' )\leq \payout_i(\Cash)$
for all strategy profiles $\Cash'$ that differ from $\Cash$ only in the move of player $i$,  
that is, for which $\Cash' = \Cash \cup \{i\}$ or $\Cash'  = \Cash \setminus \{i\}$. 
We call these simply ``Nash equilibria" henceforth, since we do not consider the probabilistic notion of 
mixed Nash equilibrium in this paper. 

Plainly, an optimum, if one exists, is also a Nash equilibrium. We may therefore 
approach the question of the existence of optima via an analysis of the Nash equilibria.  
A strategy profile that is not a Nash equilibrium is unstable, in the sense that some player has an incentive
to change their move, possibly to the detriment of another player. Where an optimum does not exist, 
it is therefore preferable that the ``solution'' to the game be a Nash equilibrium. However, multiple Nash equilibria 
may exist. A reasonable criterion for choice amongst Nash equilibria is to choose one that is optimal amongst the 
Nash equilibria: that is, a Nash equilibrium $\Cash$ such that $\Cash' \nashord \Cash$ for all Nash equilibria $\Cash'$. 
Again, there is no \emph{a priori} guarantee that such an optimum exists, however.  (We give an example of this below.)

In our application to Liquidity Event games, it is desirable that an optimum Nash equilibrium exists for all values of $V$,
since this means that there is always a distribution to the holders of the convertible instruments that
avoids inherent conflicts between investors. The contracts in question could be argued to have been 
poorly designed or poorly selected if there  were values of $V$ for which the Liquidity Event game lacks an optimum Nash equilibrium, since there could then potentially eventuate a liquidity event situation 
where any Nash equilibrium selected would leave at least one investor worse off 
than in some other Nash equilibrium. The injured party would be motivated to pursue legal  action in this 
case, seeking redress by extra-contractual means. A well-designed contract would have given greater legal certainty 
and reduce legal costs by ensuring that this possibility could not arise.  

It is plausible that the investors would  collaborate in choosing their moves so as to select an optimum Nash equilibrium when one exists, since it gives them their maximum payout amongst all stable alternatives. 
Such an agreement, once reached, also presents little risk that one of the parties to the agreement would 
maliciously change their move at the last minute. 

The following result identifies one condition under which an optimum Nash equilibrium is guaranteed to exist: when the 
total amount $V$ to be paid out is less than the total  cashout amounts due to the players. 
Roughly, this means that an investor who elects to convert rather than cash out 
runs the risk that the funds remaining to be distributed for the lower priority conversion amounts will be zero. 
This makes it a safer strategy to cash out. 

For a set $X\subseteq \Safes$, write $\principal(X)$ for $\sum_{i\in X} \principal_i$, that is, the total principal for investors in the set $X$.

\begin{thm} 
If $\Value < \principal(\Safes)$ then $\Cash = \Safes$ is the unique  Nash equilibrium. 
\end{thm} 

\begin{proof} 
Suppose that $\Value < \principal(\Safes) $. 
To show that  the strategy profile $\Cash = \Safes$ in which all investors take cash  
is the unique Nash equilibrium, suppose that $\Cash \neq \Safes$ is a Nash equilibrium. 
Note first that we must have $\principal(\Cash) < V$.
For, if $\principal(\Cash) \geq V$ then, with $i \in \overline{\Cash}$, we have 
$\payout_i(\Cash) = 0$, because all value is paid out to investors that cash out, 
but $\payout_i(\Cash \cup \{i\})>0$ since in $\Cash\cup \{i\}$, investor $i$ participates in a 
(pro-rated) cashout. This contradicts the assumption that $\Cash$ is a Nash equilibrium.    

We next argue that we cannot have $i \in \overline{\Cash}$ with $\principal(\Cash) + \principal_i \geq \Value$. 
For, suppose that this is the case. Then we would have 
$$\payout_i(\Cash \cup \{i\}) = \frac{\principal_i}{\principal(\Cash) + \principal_i}\cdot \Value~.$$ 
Also, as argued above, we have $\principal(\Cash) < \Value$, 
so $\payout_i(\Cash) = \frac{\principal_i}{\gamma_i f(\overline{\Cash})}(\Value - \principal(\Cash))$. 
From the assumption that  $\Cash$  is a Nash equilibrium, 
we have 
$$ \frac{\principal_i}{\principal(\Cash) + \principal_i}\cdot \Value \leq \frac{\principal_i}{\gamma_i f(\overline{\Cash})}(\Value - \principal(\Cash)~.$$ 
From this, we get that 
$$
\begin{array}{ll} 
(\principal(\Cash) + \principal_i - \gamma_i f(\overline{\Cash}))  \Value  & \geq   
\principal(\Cash)(\principal(\Cash)+ \principal_i) \\[5pt] 
 &  \geq  \principal(\Cash) \Value 
 \end{array} 
 $$
 by using the assumption that $\principal(\Cash) + \principal_i \geq \Value$. 
 Since $\Value >0$ it then follows that $ \principal_i\geq \gamma_i f(\overline{\Cash})$, which contradicts 
 the condition~(\ref{eq:game_well-defined-i}) for the game to be well-defined. 
 This shows that we must have $\principal(\Cash) + \principal_i < \Value$ for all $i \in \overline{\Cash}$. 
 
 In this case we have, for $i \in \overline{\Cash}$, that 
 $\payout_i(\Cash\cup\{i\}) = \principal_i$ and 
  $$\payout_i(\Cash) = \frac{\principal_i}{\gamma_i f(\overline{\Cash})}( \Value - \principal(\Cash))~. $$ 
  Since $\Cash$ is a Nash equilibrium, we have 
  $$\principal_i \leq \frac{\principal_i}{\gamma_i f(\overline{\Cash})}( \Value - \principal(\Cash))~. $$ 
  Summing over $ i \in \overline{\Cash}$, 
  and using condition~(\ref{eq:game-well-defined})
    we get 
  $$ 
  \begin{array}{ll} 
  \principal(\overline{\Cash})  = \sum_{i\in \overline{\Cash}} ~\principal_i   & \leq (\sum_{i\in \overline{\Cash}} \frac{\principal_i}{\gamma_i f(\overline{\Cash})})( \Value - \principal(\Cash)) \\[5pt] 
  & <  \Value - \principal(\Cash)
  \end{array}
 $$ 
 where, for the last inequality, we use the fact $V- \principal(\Cash) >0$, established above. 
 
But this implies that $\principal(\Safes) = \principal(\overline{\Cash}) + \principal(\Cash) < \Value$, contradicting the assumption. 
This shows that the only possible Nash equilibrium in case $\Value < \principal(\Safes) $ is $\Cash = \Safes$. 
Indeed the arguments above also show that $\Cash = \Safes$ is a Nash equilibrium, because it has been shown in the case of every strategy profile $\Cash' = \Safes\setminus \{i\}$, that $i$ gains by switching their move from conversion to cashout. 
\end{proof}

In the case where $\Value \geq \principal(\Safes)$, there are enough funds for each investor who chooses to cash out 
to receive the full amount of their principal, independently of the choices of other investors. The 
main question here, for each investor, is whether they could receive more than their principal by choosing to convert. 
The answer to that question potentially depends on the choices of other investors, since the Cashout Amounts have 
priority over the funds to be distributed to investors who convert. 

Indeed, in some instances of the Liquidity Event game, there is an inherent conflict between the interests of different investors.

\begin{example}
Consider the symmetric Liquidity Event game, with $\beta = 1 $ and $V = 8$,  and two investors with
$\principal_i = 1$, $\alpha_i =  1$, $\gamma_i = 3$ for each investor $i=1,2$. 
This gives the game depicted in Figure~\ref{fig:conflict}. (Note that the constraint~(\ref{eq:game-well-defined}) is satisfied for this game.)  The game is similar to the 
Game of Chicken \cite{RC66} in that there are two Nash equilibria (Cash,Convert) and (Convert,Cash), each with one party gaining a better payoff than the other, but if both play Convert, then both receive a lesser payoff than if they both play Cash. We note that this game has $\principal_i < \gamma_i\alpha_i$ for all $i$, so condition~(\ref{eq:solvable}) is not satisfied. 
\end{example} 

\begin{figure} 
\begin{center} 
\begin{tikzpicture}

\matrix[matrix of math nodes,every odd row/.style={align=right},every even row/.style={align=left},every node/.style={text width=1.5cm},row sep=0.2cm,column sep=0.2cm] (m) {

1&$1\frac{1}{6}$\\
1&1\\

1&$\frac{8}{9}$\\

$1\frac{1}{6}$&$\frac{8}{9}$\\
};
\draw (m.north east) rectangle (m.south west);
\draw (m.north) -- (m.south);
\draw (m.east) -- (m.west);

\coordinate (a) at ($(m.north west)!0.25!(m.north east)$);
\coordinate (b) at ($(m.north west)!0.75!(m.north east)$);
\node[above=5pt of a,anchor=base] {Cash};
\node[above=5pt of b,anchor=base] {Convert};

\coordinate (c) at ($(m.north west)!0.25!(m.south west)$);
\coordinate (d) at ($(m.north west)!0.75!(m.south west)$);
\node[left=2pt of c,text width=1cm]  {Cash};
\node[left=2pt of d,text width=1cm]  {Convert};

\node[above=18pt of m.north] (firm b) {Investor 2};
\node[left=1.6cm of m.west,rotate=90,align=center,anchor=center] {Investor 1};

\end{tikzpicture}
\end{center} 
\caption{A liquidity game with conflicting equilibria\label{fig:conflict}} 
\end{figure}

We now show that condition~(\ref{eq:solvable}) 
ensures that conflicts of this kind do not occur.
Suppose  that condition~(\ref{eq:solvable}) is satisfied.
If $\Cash$ is a Nash equilibrium then for all $i \in \Cash$, 
we have $$\principal_i  = \payout_i(\Cash) \geq
 \payout_i(\Cash\setminus \{i\})) = 
 \frac{\principal_i}{\gamma_i f(\overline{\Cash} \cup \{i\})}
(V- \principal(\Cash\setminus \{i\})) 
$$ 
or, equivalently, 
\begin{equation} 
\principal(\Cash) - \principal_i  + \gamma_i f(\overline{\Cash}) + \gamma_i \alpha_i \geq V \label{eq:nashcash} ~.
\end{equation} 
Additionally, for 
$i \in \overline{\Cash}$, 
we have 
$$\payout_i(\overline{\Cash}\cup \{i\}) = \principal_i \leq \frac{\principal_i}{\gamma_i f(\overline{\Cash})} (V- \principal(\Cash)) = \payout_i(\Cash) $$ 
which is equivalent to 
\begin{equation} 
V  \geq \principal(\Cash) + \gamma_i f(\overline{\Cash})
\label{eq:nashconv} ~.
\end{equation} 
Conversely, if these two conditions are satisfied, then $\Cash$ is a Nash equilibrium. 

Let $g: \pow{\Safes} \rightarrow \Reals$ be defined by 
$$ g(K) = \frac{V- \principal(K)}{f(\overline{K})}$$
for $K \subseteq \Safes$.   Note that for 
$i\in \overline{K}$ we have $\payout_i(K) = \principal_i g(K)/\gamma_i$. 

\begin{thm} 
\label{thm:optnash}
Assume that  $\Value \geq \principal(\Safes)$ and condition~(\ref{eq:solvable}) is satisfied. 
Suppose that there exists a Nash equilibrium, 
and let $K^*$ be a Nash equilibrium that satisfies $g(K^*) \geq g(K)$ for all 
Nash equilibria $K$. Then $K^*$ is an optimum Nash equilibrium with respect to $\nashord$. 
\end{thm}

\begin{proof} 
Assume that $\Value \geq \principal(\Safes)$ and that there exists a Nash equilibrium. 
Let $K^*\subseteq \Safes$ be a Nash equilibrium that satisfies $g(K^*) \geq g(K)$ for all 
Nash equilibria $K$.  We show that $K \nashord K^*$ for all Nash equilibria $K$. 
Let $K$ be any Nash equilibrium.  We show  
that $\payout_i(K) \leq \payout_i(K^*)$ for all $i \in \Safes$. 
We consider three cases: $i\in K$, $i \in \overline{K^*} \setminus K$ and 
$i \in K^* \setminus K$.

Since $\Value \geq \principal(\Safes)$, we have $\payout_i(K) \geq \principal_i$ for all $i\in \Safes$. 
It is immediate that for $i \in K$ we have $\payout_i(K) = \principal_i \leq \payout_i(K^*)$. 

For $i$ in neither $K^*$ nor $K$, we have, by definition of $K^*$ and the fact that $K$ is a Nash equilibrium, that 
$$\payout_i(K^*) = \principal_i g(K^*)/\gamma_i \geq \principal_i g(K)/\gamma_i = \payout_i(K) ~.$$

The remaining case is $i \in K^* \setminus K$. Here, since $i \in K^*$ and $K^*$ is a Nash equilibrium, 
by~(\ref{eq:nashcash}) we have 
$V- \principal(K^*) \leq \gamma_i f(\overline{K^*}) - \principal_i + \gamma_i\alpha_i$. 
Thus 
$$\frac{g(K^*)}{\gamma_i} = \frac{V- \principal(K^*) }{ \gamma_if(\overline{K^*})} 
\leq 1 - \frac{(\principal_i - \gamma_i\alpha_i)}{ \gamma_if(\overline{K^*})}~.$$ 
By condition~(\ref{eq:solvable}),  
$\principal_i - \gamma_i\alpha_i \geq 0$, so we have $\frac{g(K^*)}{\gamma_i} \leq 1$. 
Thus, from the definition of $K^*$ and the fact that $K$ is a Nash equilibrium, 
$$ \payout_i(K) = \frac{\principal_i g(K)}{\gamma_i} \leq \frac{\principal_i g(K^*)}{\gamma_i} \leq \principal_i = 
\payout_i(K^*)~.$$
\end{proof}

\begin{thm} \label{thm:gamma-ord}
Suppose  that  $\Value \geq \principal(\Safes)$ and 
condition~(\ref{eq:solvable}) is satisfied.
If $\Cash$ is a Nash equilibrium, then for all $i\in \Cash$ and $j \in \overline{\Cash}$ we have 
$\gamma_i \geq \gamma_j$. Moreover, if $\gamma_i = \gamma_j$ then $\principal_i = \gamma_i\alpha_i$ and 
$g(\Cash) = \gamma_i$.
\end{thm}

\begin{proof} 
If $i \in \Cash$ then we have $V\leq \principal(\Cash) + \gamma_i f(\overline{\Cash}) + \gamma_i \alpha_i - \principal_i$
by~(\ref{eq:nashcash}).
If $j \in \overline{\Cash}$ then we have $\principal(\Cash) + \gamma_jf(\overline{\Cash}) \leq V$
by~(\ref{eq:nashconv}). 
Combining the two inequalities, we derive 
$(\gamma_i - \gamma_j)f(\overline{\Cash}) \geq \principal_i - \gamma_i \alpha_i$. 
By condition~(\ref{eq:solvable}), the term on the right hand side  
of this inequality is non-negative and $f(\overline{\Cash})$ is positive, so it follows that $\gamma_i \geq \gamma_j$. 
Moreover if  $\gamma_i = \gamma_j$ we have 
$0 \geq \principal_i - \gamma_i \alpha_i \geq 0$, so $\principal_i = \gamma_i\alpha_i$ and 
our inequalities state 
$\principal(\Cash) + \gamma_i f(\overline{\Cash})  \leq V\leq \principal(\Cash) + \gamma_i f(\overline{\Cash})  $, so 
$V = \principal(\Cash) + \gamma_i f(\overline{\Cash})$. It follows from this that 
$g(\Cash) = \gamma_i$.
\end{proof} 

We see from this result that if condition~(\ref{eq:solvable}) is satisfied, 
we can restrict the sets $\Cash$ that need to be considered to identify the Nash equilibria 
to those sets such that for some $i \in \Safes$ we have $j \in \overline{K}$ implies $\gamma_j \geq \gamma_i$. 
This observation will be useful for obtaining efficient algorithms below. 
Indeed a smaller set of cases suffices. The following result shows that if there is a Nash equilibrium in which some
$\gamma_i$ occurs for both $\Cash$ and $\overline{\Cash}$, then there is an $\approx$-equivalent Nash equilibrium 
that does not have such a ``boundary crossing''.

\begin{thm} \label{thm:nash-shift} 
Assume that  $\Value \geq \principal(\Safes)$  and condition~(\ref{eq:solvable}) is satisfied. 
Fix $i \in \Safes$ and let   $E =   \{k\in \Safes~|~\gamma_k = \gamma_i\}$, and  
$G =   \{k\in \Safes~|~\gamma_k > \gamma_i\}$. 
Suppose that  $Y \subseteq E$.  
If $Y \neq \emptyset$ and $E\setminus Y \neq \emptyset$
and $G\cup Y$ is a Nash equilibrium, then 
\begin{itemize} 
\item[(i)] $\principal_k = \gamma_k\alpha_k$ for all $k\in Y$ 
\item[(ii)] $g(G) = \gamma_i$
\item[(iii)] $G$ is a Nash equilibrium
\end{itemize} 
Conversely, if (i)-(iii) then $G\cup Y$ is a Nash equilibrium. 
Moreover (i) and (ii) imply $g(G) = g(G\cup Y)$ and $G \approx G \cup Y$.  
\end{thm} 

\begin{proof} 
Let  $L = \{k\in \Safes~|~\gamma_k < \gamma_i\}$.

We begin by proving the final part of the result. We first show that 
  (i) and (ii) imply that $g(G) = g(G\cup Y)$. 
Note that 
\begin{align*} 
 \principal(G\cup Y ) & + \gamma_i f(\overline{G \cup Y})  \\
 & =  \principal(G) + \principal(Y) + \gamma_i f(\overline{G})  - \gamma_i \sum_{k \in Y} \alpha_k\\ 
  & =  \principal(G) + \gamma_i f(\overline{G})  + \sum_{k \in Y} (\principal_k -\gamma_k \alpha_k) & \text{(by $\gamma_i=\gamma_k$ for $k\in Y$)}\\ 
 & =  \principal(G) + \gamma_i f(\overline{G})  & \text{(by (i))} \\
 & = V & \text{(by (ii))} 
\end{align*} 
It follows that $g(G\cup Y) = \gamma_i = g(G)$. 

Next, we show that if $ g(G)= g(G\cup Y) $  and (ii) then $G\cup Y \approx G$.
For this, we consider three cases $k \in G$, $k \in Y$ and $k \in \overline{G\cup Y}$ and 
show $\payout_k(G\cup Y) = \payout_k(G)$ in each. 
\begin{itemize} 
\item For $k \in G$, we have also $k \in G \cup Y$, so $\payout_k(G \cup Y) = \principal_k = \payout_k(G)$,  by definition. 
\item For $k\in Y$, we have $k \in E$ so $\gamma_k = \gamma_i$. 
Also, $k \not \in G$. Thus 
 \begin{align*} 
 \payout_k(G\cup Y) & =   \principal_k\\
 & =  \principal_k g(G)/\gamma_i & \text{(by (ii))} \\ 
 & =   \principal_k g(G)/\gamma_k \\
 & = \payout_k(G)~.
 \end{align*}
\item 
Finally, for $k \in \overline{G\cup Y}$, we have $k \in \overline{G}$,  so 
$ \payout_k(G\cup Y) = \principal_k g(G\cup Y)/\gamma_k =   \principal_k g(G)/\gamma_k = \payout_k(G)$.
\end{itemize}  
 
Suppose now that $Y \subset E$ with $Y \neq \emptyset$ and $E\setminus Y \neq \emptyset$
and $G\cup Y$ is a Nash equilibrium. 
Claims (i) and (ii) follow by Theorem~\ref{thm:gamma-ord}. We need to show (iii), that $G$ is a Nash equilibrium. 

First, we need to show that 
\begin{equation} 
\principal(G) - \principal_k  + \gamma_k f(\overline{G}) + \gamma_k \alpha_k \geq V 
\label{eq:Gnash} 
\end{equation} 
for all $k \in G$. Let $k \in G$. Then $k \in G\cup Y$ and $\gamma_k > \gamma_i$. 
Since $G\cup Y$ is a Nash equilibrium and $k \in G\cup Y$, we have 
$$\principal(G\cup Y) - \principal_k  + \gamma_k f(\overline{G\cup Y}) + \gamma_k \alpha_k \geq V $$
by~(\ref{eq:nashcash}). Equivalently, 
$$(\principal(G) - \principal_k  + \gamma_k f(\overline{G}) + \gamma_k \alpha_k) + 
(\principal(Y) - \gamma_k \sum_{j\in Y} \alpha_j)  \geq V $$
The desired conclusion~(\ref{eq:Gnash}) now follows if 
$\principal(Y) - \gamma_k \sum_{j\in Y} \alpha_j \leq 0$. 
But this indeed holds, since we have, for all $j \in Y$, that  $\principal_j = \gamma_j \alpha_j = 
\gamma_i \alpha_j < \gamma_k \alpha_j$.  

Next, we need to show that $V  \geq \principal(G) + \gamma_k f(\overline{G})$ for all $k \in \overline{G}$. 
Now, $\overline{G}= \overline{G\cup Y} \cup Y$. 
Since $G\cup Y$ is a Nash equilibrium, we have  by~(\ref{eq:nashconv}) that 
$\gamma_k \leq g(G\cup Y)$ for $k \in \overline{G\cup Y}$. Since $g(G\cup Y) = g(G)$, 
it follows that $\gamma_k \leq g(G)$ for $k \in \overline{G\cup Y}$, which gives the desired inequality in this case. 
For $k \in Y$, the desired inequality is immediate (and holds with equality)  from the fact (ii) that $g(G) = \gamma_i$. 
This completes the proof that $G$ is a Nash equilibrium. 
\end{proof} 

We see from this result that to identify the inequivalent Nash equilibria, it suffices to consider just 
the linear number of sets $\Cash_{\gamma}= \{k\in \Safes~|~\gamma_k \geq  \gamma\}$
where $\gamma = \infty$ or $\gamma = \gamma_i$ for some $i \in \Safes$. 
Other strategy profiles may be Nash equilibria, but if, so, they are $\approx$-equivalent to one of these profiles. 

This result directly yields a polynomial time algorithm for deciding the existence of a Nash equilibrium and, in the the case that 
Nash equilibria exist, computing a representative for each $\approx$-equivalence class of the Nash equilibria, 
as well as an optimum Nash equilibrium. 
For each $\Cash = \Cash_\gamma$ where $\gamma =\infty $ or $\gamma = \gamma_i$ for $i \in \Safes$, 
and for each $k\in \Safes$, we compute the 
number $\principal(\Cash) + \gamma_k f(\overline{\Cash})$
and check whether  $\principal(\Cash) + \gamma_k f(\overline{\Cash}) \geq V- \principal_k + \gamma_k \alpha_k$
(if $k \in \Cash)$ and 
$\principal(\Cash) + \gamma_k f(\overline{\Cash}) \leq V$ (if $k \in \overline{\Cash}$. 
Any instance $\Cash$ satisfying these conditions is a Nash equilibrium, and if none does, 
then there are no Nash equilibria.  The optimum Nash equilibrium can be identified as the 
one that maximizes $g(\Cash_\gamma)$, by Theorem~\ref{thm:optnash}.

The time complexity of the algorithm is $O(|\Safes| \cdot (|\Safes| + m(b)))$, 
where $b$ is the bitlength of the numbers defining the problem 
and $m(b)$ is the cost of multiplying two $b$-bit numbers.\footnote{
Theoretically, the bound  $m(b) = O(b\,\log(b))$ is known \cite{HJ21}, but asymptotically less efficient algorithms may be preferable in practice.} We discuss some special cases  below where this computation can be optimized to give a lower complexity.

We will later (see Theorem~\ref{thm:nash-shift2}) identify a condition, strengthening 
condition~(\ref{eq:solvable}), 
under which we have a  converse to Theorem~\ref{thm:nash-shift}: 
if $G$ is a Nash equilibrium with $g(G) = \gamma_i$, then for all $Y\subseteq E$, we have that $G\cup Y$ is a Nash equilibrium. 
However, this converse does not follow just from condition~(\ref{eq:solvable}), as the following example shows. 
This means that while we have an algorithm that computes a complete set of representatives of the Nash equilibria, 
we do not yet have a concrete complete representation of the set of all Nash equilibria, and still need to 
evaluate each of the (potentially exponentially many) sets $G \cup Y$ individually.  

\begin{example} \label{ex:notYconverse} 
Consider $\Safes = \{0,1,2\}$, with parameters as given  in the following table 
$$ 
\begin{array}{r|ccc} 
i & 0 & 1 & 2 \\ 
\hline 
\alpha_i & 1 & 2 & 1 \\ 
\gamma_i & 3 & 2 & 2 \\ 
\principal_i  & 5\frac{1}{2} & 4 & 2
\end{array} 
$$
Let $\beta = 1 $ and $V = 13 \frac{1}{2}$. 
Note that here we have $\gamma_1= \gamma_2 < \gamma_0$, so, with $i=1$, 
we have $G = \{0\}$ and $E = \{1,2\}$. The following can be easily verified
using the characterization of Nash equilibria given by conditions (\ref{eq:nashcash}) and (\ref{eq:nashconv}): 
\begin{itemize} 
\item The game satisfies constraint~(\ref{eq:game-well-defined}).  
\item $G$ is Nash equilibrium, and $g(G) = \gamma_1$. 
\item $G \cup \{2\}$ is a Nash equilibrium. This means with $Y = \{2\}\subseteq E$,  we have a ``boundary crossing'' $2 \in  G \cup Y$, 
$1 \in  \overline{G \cup Y}$ with $\gamma_1 = \gamma_2$, as in the assumptions of Theorem~\ref{thm:nash-shift}. 
\item $G \cup \{1\}$ is not a Nash equilibrium, because  (\ref{eq:nashcash}) is not satisfied for player 0. 
\end{itemize} 
\end{example} 

However, while it guarantees that there exist an optimum Nash equilibrium, if any, 
condition~(\ref{eq:solvable}) is not sufficient to guarantee  
that there is a Nash Equilibrium, as shown by the following example.  
We identify some sufficient conditions for the existence of Nash equilibria in the following sections. 

\begin{example}\label{ex:noeq}
Consider the Liquidity Event game, with $\beta = 1 $ and $V = 29$,  and two investors 1,2 with
$\principal_1 = 10$, $\alpha_1 =  6$, $\gamma_1 = 1$,  
$\principal_2= 16$, $\alpha_2 =  5$, $\gamma_2 = 3$. 
(Again the constraint~(\ref{eq:game-well-defined}) is satisfied with these parameters.) 
This gives the game depicted in Figure~\ref{fig:nonash}.  In this case, there is a cycle of 
best-reponses 2:Cash $\rightarrow$ 1:Convert $\rightarrow$ 2:Convert $\rightarrow$ 1:Cash $\rightarrow$  2:Cash, 
where $i:a \rightarrow j:b$ denotes that $j$ playing $b$ is a best response to  $i$ playing $a$. 
Thus, this game has no Nash equilibrium.  
Note that this game has $\principal_1 > \gamma_1 \alpha_1$ and $\principal_2 > \gamma_2 \alpha_2$, 
so condition~(\ref{eq:solvable}) is satisfied. 
\end{example} 

\begin{figure} 
\begin{center} 
\begin{tikzpicture}

\matrix[matrix of math nodes,every odd row/.style={align=right},every even row/.style={align=left},every node/.style={text width=1.5cm},row sep=0.2cm,column sep=0.2cm] (m) {
16&$15\frac{1}{5}$\\

10&10\\

16&$18\frac{14}{25}$\\

$10\frac{5}{6}$&$9\frac{2}{3}$\\
};
\draw (m.north east) rectangle (m.south west);
\draw (m.north) -- (m.south);
\draw (m.east) -- (m.west);

\coordinate (a) at ($(m.north west)!0.25!(m.north east)$);
\coordinate (b) at ($(m.north west)!0.75!(m.north east)$);
\node[above=5pt of a,anchor=base] {Cash};
\node[above=5pt of b,anchor=base] {Convert};

\coordinate (c) at ($(m.north west)!0.25!(m.south west)$);
\coordinate (d) at ($(m.north west)!0.75!(m.south west)$);
\node[left=2pt of c,text width=1cm]  {Cash};
\node[left=2pt of d,text width=1cm]  {Convert};

\node[above=18pt of m.north] (firm b) {Investor 2};
\node[left=1.6cm of m.west,rotate=90,align=center,anchor=center] {Investor 1};

\end{tikzpicture}
\end{center} 
\caption{A liquidity game with no Nash equilibrium\label{fig:nonash}} 
\end{figure}

\section{Special Case: Pre-Money SAFEs} \label{sec:nash-pre-money} 

In this section we consider a special case, 
where  $\Value \geq \principal(\Safes)$ and  $\principal_i = \gamma_i\alpha_i$ for all $i\in \Safes$. 
As discussed in Sections~\ref{sec:safe-cap},~\ref{sec:discount} and~\ref{sec:cap-discount}, 
all the nontrivial Pre-Money SAFEs variants  satisfy this condition, 
as  does the Post-Money SAFE with Discount Only of Section~\ref{sec:postdiscount}. 

We show that Nash equilibria are guaranteed to exist in this case.  
Condition~(\ref{eq:solvable}) is satisfied under this assumption so, 
by Theorem~\ref{thm:optnash},  an optimum Nash equilibrium is 
also guaranteed to exist in this case. We also show that an improvement 
of the algorithm for computing Nash equilibria is possible. 

In this case, the conditions for $\Cash$ to be a Nash equilibrium simplify to the following. 
For $i \in \Cash$,  
\begin{equation} 
\principal(\Cash) + \gamma_i f(\overline{\Cash}) \geq V \label{eq:nashcashPre}
\end{equation} 
and for $i \in \overline{\Cash}$,  
\begin{equation} 
V  \geq \principal(\Cash) + \gamma_i f(\overline{\Cash})
\label{eq:nashconvPre} ~.
\end{equation} 
Equivalently, since $f(\Cash)>0$, we have that $\Cash$ is a Nash equilibrium when 
\begin{equation} 
\principal(\Cash) + \min(\{\gamma_i ~|~i \in \Cash\})f(\overline{\Cash}) \geq V \label{eq:nashcashPremin}
\end{equation} 
and 
\begin{equation} 
V \geq \principal(\Cash) + \max(\{\gamma_i ~|~i \in \overline{\Cash}\})f(\overline{\Cash})  ~. \label{eq:nashcashPremax}
\end{equation} 
where we treat the cases of empty sets by $\min (\emptyset) = \infty$ and $\max(\emptyset) = -\infty$   
so that, respectively,  the  constraint
(\ref{eq:nashcashPremin}) or  (\ref{eq:nashcashPremax}) is trivial when $\Cash = \emptyset$ or 
$\overline{\Cash} = \emptyset$. 

From Theorem~\ref{thm:gamma-ord}, we have that if $\Cash$ is a Nash equilibrium and 
 $i\in \Cash$ and $j \in \overline{\Cash}$ then  $\gamma_i \geq \gamma_j$.
 Hence $\max(\{\gamma_i ~|~i \in \overline{\Cash}\}) \leq \min(\{\gamma_i ~|~i \in \Cash\})$, 
 and  we have that 
 $\Cash$ is a Nash equilibrium just when $V$ is in the interval 
 \begin{equation}
 [~\principal(\Cash) + \max(\{\gamma_i ~|~i \in \overline{\Cash}\})\cdot f(\overline{\Cash})~,~
 \principal(\Cash) + \min(\{\gamma_i ~|~i \in \Cash\})\cdot f(\overline{\Cash})~] 
\label{eq:interval}
 \end{equation}

In Theorem~\ref{thm:nash-shift} we saw that if $K$ is a Nash equilibrium and there exists $i\in K$ and $j\in \overline{K}$ 
with $\gamma_i = \gamma_j$, then  
 $G = \{k \in \Safes~|~ \gamma_k \geq \gamma_i\}$ is also a Nash equilibrium, $g(K) = \gamma_i$,  
and $G \approx K$. 
The following result shows that in the present special case, we have converse to this result: for 
any $Y \subseteq \{k \in \Safes~|~\gamma_k = \gamma_i\}$, we have that $G \cup Y$ is a Nash equilibrium. 
This yields a complete characterization of the (pure strategy) Nash equilibria in this special case.
(In general, there may be $G\cup Y$ that are not Nash equilibria, and it is necessary to test each individually.)

\begin{thm} \label{thm:nash-shift2} 
Suppose that   $\Value \geq \principal(\Safes)$  and 
$\principal_k = \gamma_k\alpha_k$ for all $k\in \Safes$. 
Fix $i \in \Safes$. Let $E =   \{k\in \Safes~|~\gamma_k = \gamma_i\}$, and let 
$G =   \{k\in \Safes~|~\gamma_k > \gamma_i\}$. 
Suppose $G$ is a Nash equilibrium and $g(G) = \gamma_i$. 
Then for all $Y\subseteq E$, we have that $G\cup Y$ is a Nash equilibrium and 
$G\cup Y \approx \Cash$. Moreover,  $\payout_k(G\cup Y) = \principal_k$ for all $k \in E$.
\end{thm} 

\begin{proof} 
Suppose that $G$ is a Nash equilibrium with $g(G) = \gamma_i$ and $Y\subseteq E$. 
We show that $G\cup Y$ is a Nash equilibrium. We satisfy conditions (i) and (ii) of Theorem~\ref{thm:nash-shift}, 
so we have  $g(G) = g(G\cup Y)$ and $G \approx G\cup Y$. 
Since $\principal_k = \gamma_k\alpha_k$ for all $k\in \Safes$, 
we have from (\ref{eq:nashcashPre}) and (\ref{eq:nashconvPre}) 
that $K$ is a Nash equilibrium iff $\gamma_k \geq g(K)$ for $k \in K$ and 
$g(K) \geq \gamma_k $ for $k \in \overline{K}$.  
It is  immediate from the definitions that these conditions hold
for $K=G\cup Y$, so this is a Nash equilibrium. 
\end{proof} 

In this special case, we obtain an improvement of the algorithm for identifying the Nash equilibria up to $\approx$-equivalence. 
As before, it suffices to consider just 
the linear number of sets $\Cash_{\gamma}= \{k\in \Safes~|~\gamma_k \geq  \gamma\}$
where $\gamma = \infty$ or $\gamma = \gamma_i$ for some $i \in \Safes$. 
However, the test that we perform for each value of $\gamma$ can be optimized.
Rather than performing a test for each value of $\gamma$ and each $k \in \Safes$, 
we can now just compute the interval~(\ref{eq:interval}) for each value of $\gamma$,
and determine whether it contains $V$. We note that, after sorting the $\gamma_i$,
each of the terms in the endpoint values of these intervals
can be incrementally computed from the corresponding terms in the preceding interval 
as we scan through $\Safes$ in order of increasing $\gamma_i$, 
adding or subtracting appropriate values $\principal_k$ or $\alpha_k$ 
and adjusting the maxima and minima as we go. 
As before, an optimum Nash equilibrium can be identified as the 
one that maximizes $g(\Cash_\gamma)$, by Theorem~\ref{thm:optnash}.

The complexity of this variant of the algorithm is $O(|\Safes| \cdot (\log |\Safes| + m(b)))$, 
where $b$ is the bitlength of the numbers defining the problem 
and $m(b)$ is the cost of multiplying two $b$-bit numbers. 

As above, other strategy profiles may be Nash equilibria, but if, so, they are $\approx$-equivalent to one of 
the strategy profiles considered by the algorithm. However, in this special case,
Theorem~\ref{thm:nash-shift2} also provides a straightforward way to produce \emph{all} the Nash equilibria
from the ones identified by the algorithm. (We are guaranteed that $G\cup Y$ 
is a Nash equilibrium whenever $G$ is a Nash equilibrium with $g(G) = \gamma_i$, in this case.) 

Moreover, the following result shows that, indeed, one of the  strategy profiles considered by the algorithm is a Nash equilibrium. 

\begin{thm} \label{thm:existsNash1}
Suppose that  $\Value \geq \principal(\Safes)$ and $\principal_i = \gamma_i\alpha_i$ for all $i\in \Safes$. 
Then there exists a Nash equilibrium.
\end{thm} 

\begin{proof} 
We assume that there does not exist a Nash equilibrium, and derive a contradiction. 
Consider the sets $\Cash_\gamma = \{ k\in \Safes~|~\gamma \leq \gamma_k \}$, 
where either $\gamma = \gamma_i$ for some $i \in \Safes$ or $\gamma = \infty$. 
Note that $\Cash_{\infty} = \emptyset$ and for the least $\gamma_i$, we have 
$\Cash_{\gamma_i} = \Safes$. Also, $\overline{\Cash_{\gamma}} = \{ k\in \Safes~|~ \gamma_k < \gamma\}$. 

For $i \in \Safes$, let $\gamma_i^+$ be the least value $\gamma_j > \gamma_i$ for $j \in \Safes$, 
else $\infty$ if  there is no such value. 
Then we have that 
$$\min(\{\gamma_k ~|~ k \in \Cash_{\gamma_i}\}) = \gamma_i = 
\max(\{\gamma_k ~|~ \gamma_k < \gamma_i^+, ~ k \in \Safes\}) = \max (\{\gamma_k ~|~ k \in \overline{\Cash_{\gamma^+_i}}\})$$
Hence 
\begin{align*} 
~~& \principal(\Cash_{\gamma_i})  ~+~  \min(\{\gamma_k ~|~ k \in \Cash_{\gamma_i}\}) \cdot  f(\overline{\Cash_{\gamma_i}})\\[10pt] 
&  = \sum_{k\in \Safes, \gamma_i \leq \gamma_k} \principal_k 
 ~+~ \gamma_i \cdot ( \beta ~+ \sum_{k\in \Safes,~\gamma_k < \gamma_i} \alpha_k)\\[10pt]
 &  = 
 \sum_{k\in \Safes, \gamma_i^+ \leq \gamma_k} \principal_k
 +~ \sum_{k\in \Safes, \gamma_i = \gamma_k} \principal_k
 ~+~ \gamma_i \cdot ( \beta ~+ \sum_{k\in \Safes,~\gamma_k < \gamma_i} \alpha_k)\\[10pt]
  &  = 
 \sum_{k\in \Safes, \gamma_i^+ \leq \gamma_k} \principal_k
 ~+~ \sum_{k\in \Safes, \gamma_i = \gamma_k} \gamma_k\alpha_k
 ~+~ \gamma_i \cdot ( \beta ~+ \sum_{k\in \Safes,~\gamma_k < \gamma_i} \alpha_k)\\[10pt]
   &  = 
 \sum_{k\in \Safes, \gamma_i^+ \leq \gamma_k} \principal_k
 ~+~ \gamma_i \cdot ( \beta ~+
  \sum_{k\in \Safes, \gamma_i = \gamma_k} \alpha_k
 ~+~
  \sum_{k\in \Safes,~\gamma_k < \gamma_i} \alpha_k)\\[10pt]
 &  = 
 \sum_{k\in \Safes, \gamma_i^+ \leq \gamma_k} \principal_k
 ~+~ \gamma_i \cdot ( \beta ~+
  \sum_{k\in \Safes,  \gamma_k < \gamma_i^+ } \alpha_k)\\[10pt]
 &  = 
 \principal(\Cash_{\gamma_i^+}) 
 ~+~ \max (\{\gamma_k ~|~ k \in \overline{\Cash_{\gamma^+_i}}\})
 \cdot f(\overline{\Cash_{\gamma_i^+}})
 \end{align*} 
 That is, the  right endpoint of the interval~(\ref{eq:interval}) for $K_{\gamma_i}$ is equal to the left endpoint of the interval for 
 $K_{\gamma_i^+}$. Since the leftmost endpoint of these intervals is $-\infty$ and the right endpoint is 
 $\infty$, one of these intervals contains $V$, and there must exist a Nash equilibrium. 
\end{proof}

Note also that the (optimum) Nash equilibrium is not necessarily unique, indeed, from Theorem~\ref{thm:nash-shift2}, we see that there may be an exponential number of Nash equilibria in the worst case. 
The following example illustrates this phenomenon. 

\begin{example} 
Consider the Liquidity game with $\Safes = \{1, \ldots ,n\}$, $V= n+1$, $\beta =1$, and
$\principal_i = 1$, $\gamma_i = 1$ and $\alpha_i =1$  for each $i \in \Safes$. 
Plainly $\principal_i = \gamma_i \alpha_i$ for all $i\in \Safes$ in this case. 
In this game, all sets $\Cash \subseteq \Safes$ 
are Nash equilibria, and they all yield the same payouts $\payout_i(\Cash) = 1$ for all players $i$. 
This is obvious for $i\in \Cash$. For $i \not \in \Cash$, 
we have 
\begin{align*}
\payout_i(\Cash) 
& = \principal_i \frac{(V- \principal(K))}{\gamma_i(\beta + \sum_{i \in \overline{K}} \alpha_i)}\\
& = \frac{(n+1 - |K|)}{(1 + |\overline{K}|)} \\
& =1
\end{align*} 
since $|\overline{K}|= |\Safes| - |K| = n-|K|$. 
\end{example}

\section{Special Case: Post-Money SAFEs} \label{sec:nash-post-money}

In this section we consider the existence of Nash equilibria in the case where $\principal(\Safes) > V$
and $\alpha_i = 0 $ for all $i \in \Safes$. The Post-Money SAFE of Section~\ref{sec:post-money-cap} 
falls within this case. 
We show that Nash Equilibria exist in this case.  
Condition~(\ref{eq:solvable}) is satisfied in this case so, by Theorem~\ref{thm:optnash}, one of these equilibria is an optimum.

\begin{thm} 
Suppose that $\alpha_i = 0$ for all $i \in \Safes$. Then there exists a Nash equilibrium. 
\end{thm}

\begin{proof} 
In the case where $\alpha_i=0$ for all  $i \in \Safes$, the 
conditions for $\Cash$ to be a Nash equilibrium reduce to the following. 
\begin{equation} 
\principal(\Cash) + \min(\{\gamma_k\beta - \principal_k~|~k\in \Cash\}) \geq V \label{eq:nashcashPostmin}
\end{equation} 
and 
\begin{equation} 
V \geq \principal(\Cash) + \max(\{\gamma_k\beta ~|~k\in \overline{\Cash}\})  ~. \label{eq:nashcashPostmax}
\end{equation} 

Suppose that there is no Nash equilibrium.
Then for all $\Cash \subseteq \Safes$, we have 
$\principal(\Cash) + \min(\{ \gamma_k\beta - \principal_k~|~k \in \Cash\}) <V $ 
or 
$V < \principal(\Cash) + \max(\{ \gamma_k\beta ~|~k \in \overline{\Cash}\})$.
We derive a contradiction.

Let $\Safes = \{1, \ldots,n\}$ be sorted so that  $\gamma_1 \leq  \gamma_2 \leq \ldots \leq  \gamma_n$.  
Define $i^*$ to be the value of $i\in \Safes$ at which $\gamma_i\beta - \principal_i$ takes its minimum. 
Let $\Phi(i)$ be the following proposition: 
\begin{quote}
$i^* \neq i$ and 
 $V < \gamma_i\beta + \sum_{k= i+1}^n \principal_k$.
\end{quote} 
In the case $i=n$ we take the summation to be equal to zero. 
We show that $\Phi(i)$ holds for all $i=1 \ldots n$, by a reverse induction. 
Note that this implies that $i^* \not \in \{1, \ldots, n\}$, which is a contradiction. 

For the base case of $\Phi(n)$, we argue as follows. 
Note first that, from the fact that $\Cash = \emptyset$ is not a Nash equilibrium, we 
have that $V < \principal(\emptyset) + \max(\{ \gamma_k\beta ~|~k \in \Safes \}) = 
\gamma_n\beta$, which is the right hand conjunct of $\Phi(n)$. 
Secondly, from the fact that $\Cash = \Safes$ is not a Nash equilibrium, 
we have that  
\begin{equation} 
\principal(\Safes) + \gamma_{i^*}\beta - \principal_{i^*} <V~. \label{eq:Sistar} 
\end{equation}
If we had $i^* = n$, it would follow that 
$\principal(\Safes) + \gamma_{n}\beta - \principal_{n}  < \gamma_n\beta$, 
which is impossible since $\principal(\Safes)  - \principal_{n} \geq 0$. 
Thus, $i^* \neq  n$, and we conclude that $\Phi(n)$ holds. 

For the inductive step, suppose that $\Phi(i), \ldots, \Phi(n)$, where $i>1$. 
We show that $\Phi(i-1)$. 
Consider 
$\Cash = \{i,  \ldots, n\}$. Since this is not a Nash equilibrium, we have either  
we have that either 
\begin{equation} 
V <  \gamma_{i-1}\beta + \sum_{k=i}^n \principal_k ~~\mbox{or}~~  
V > \sum_{k=i}^n \principal_k +  \min(\{ \gamma_k\beta - \principal_k~|~i \leq k \leq n\})~. 
\label{eq:disj} 
\end{equation}

Suppose first that the second disjunct of~(\ref{eq:disj}) holds, and let 
the minimum be attained at value $k^*\in \{i, \ldots,n\}$. 
Then 
$V > \gamma_{k^*}\beta - \principal_{k^*} + \sum_{k=i}^n \principal_k $. 
However, by the instance $\Phi(k^*)$ of the  induction hypothesis, we have that 
$V < \gamma_{k^*}\beta + \sum_{k= k^*+1}^n \principal_k$. 
It follows that 
$$\gamma_{k^*}\beta - \principal_{k^*} + \sum_{k=i}^n \principal_k  < \gamma_{k^*}\beta + \sum_{k= k^*+1}^n \principal_k~.$$
But this yields that 
$$\sum_{k=i}^{k^*-1} \principal_k  < 0$$
which is an impossibility, even when the summation on the left is zero. 
We conclude that the second disjunct cannot hold. 

Thus, the left disjunct  $V <  \gamma_{i-1}\beta + \sum_{k=i}^n \principal_k $
of~(\ref{eq:disj}) holds, which is the right hand conjunct of $\Phi(i-1)$. 
If we had  $i^* = i-1$, by~(\ref{eq:Sistar}) we would have 
$$\principal(\Safes) + \gamma_{i-1}\beta -\principal_{i-1}  < V < \gamma_{i-1}\beta + \sum_{k=i}^n \principal_k$$ 
This yields 
$$ \sum_{k=1}^{i-2} \principal_k <0$$
again an impossibility. Hence $i^* \neq i-1$, and we have shown $\Phi(i-1)$. 
\end{proof} 

Since we have that condition~(\ref{eq:solvable})  is satisfied strictly in the case under consideration,
by Theorem~\ref{thm:gamma-ord}, we cannot have a Nash equilibrium 
with $i\in \Cash$ and $j \in \overline{\Cash}$ such that $\gamma_i = \gamma_j$. 
The Nash equilibria therefore lie inside the linear number of  possible cases 
$\Cash_\gamma = \{ k\in \Safes ~|~ \gamma_k\geq  \gamma\} $ where $\gamma = \infty$ or 
$\gamma = \gamma_i$ for some $i\in \Cash$.
The algorithm of Section~\ref{sec:general} therefore computes not just a set of representatives, but the set of 
all Nash equilibria.

However, it remains the case that Nash equilibria are not necessarily unique, as shown by the 
following example.

\begin{example} 
Consider the symmetric two player game with $\alpha_1= \alpha_2 = 0$, $\beta =1$, 
$\principal_1 = \principal_2 = 2$, $\gamma_1 = \gamma_2 = 5$ and $V = 6$. 
The game matrix for this game is shown in Figure~\ref{fig:nonunique}. 
There are two Nash equilibria: (Cash,Cash) with payouts $(2,2)$, 
and (Convert,Convert) with payouts $(2\frac{2}{5},2\frac{2}{5})$, so that we have  (Cash,Cash) $\nashord$(Convert,Convert). 
\end{example} 

\begin{figure} 
\begin{center} 
\begin{tikzpicture}

\matrix[matrix of math nodes,every odd row/.style={align=right},every even row/.style={align=left},every node/.style={text width=1.5cm},row sep=0.2cm,column sep=0.2cm] (m) {
2&$1\frac{3}{5}$\\

2&2\\

2&$2\frac{2}{5}$\\

$1\frac{3}{5}$&$2\frac{2}{5}$\\
};
\draw (m.north east) rectangle (m.south west);
\draw (m.north) -- (m.south);
\draw (m.east) -- (m.west);

\coordinate (a) at ($(m.north west)!0.25!(m.north east)$);
\coordinate (b) at ($(m.north west)!0.75!(m.north east)$);
\node[above=5pt of a,anchor=base] {Cash};
\node[above=5pt of b,anchor=base] {Convert};

\coordinate (c) at ($(m.north west)!0.25!(m.south west)$);
\coordinate (d) at ($(m.north west)!0.75!(m.south west)$);
\node[left=2pt of c,text width=1cm]  {Cash};
\node[left=2pt of d,text width=1cm]  {Convert};

\node[above=18pt of m.north] (firm b) {Investor 2};
\node[left=1.6cm of m.west,rotate=90,align=center,anchor=center] {Investor 1};

\end{tikzpicture}
\end{center} 
\caption{A liquidity game with two Nash equilibria\label{fig:nonunique}} 
\end{figure}

\section{Mixed SAFE Types} \label{sec:mixed} 

The analysis of the previous sections has assumed that all SAFE's issued  by the company are of the same type, that is, are all 
Pre-Money SAFEs or are all Post-Money SAFEs. We now present an example that shows that mixing different 
types of SAFEs can create situations where no pure strategy Nash equilibria exist. We consider the combination of a
Pre-Money SAFE with Cap Only, with principal $\principal_1$ and cap $\Cap_1$, and
a Post-Money SAFE with Cap Only, with principal $\principal_2$ and a Post-Money Cap $\Cap_2$.

A first issue is to understand the interaction of Pre-Money SAFEs and Post-Money SAFEs in the calculation of the number of 
shares issued in conversion. Essentially, this is done
by first determining the number of shares to be issued to the Pre-Money SAFE, and then calculating the number of shares
that should be issued to the Post-Money SAFE. 
We follow the notation from  Section~\ref{sec:post-money-cap}. Let $\Converted = \overline{\Cash}$ be the
set of SAFEs that convert. (Note that in this section, payouts $\payout_i(\Cash)$ are expressed in terms of $\Cash$, 
but numbers of shares $\sconverted{\Converted}_i$ are expressed in terms of $\Converted$.) 

We first determine the number of shares $\sconverted{\Converted}_1$ issued to the Pre-Money SAFE, in the event that this is converted. 
Recall from Section~\ref{sec:safe-cap} that the Liquidity Capitalization for
the Pre-Money SAFE excludes  SAFEs, so it can be treated as a constant, equal to the number of 
common shares $\scommon$ in our simple scenario.  We obtain that the number of shares issued in 
conversion to the Pre-Money SAFE is $\sconverted{\Converted}_1=\principal_1\scommon/\Cap_1$, exactly as in 
Section~\ref{sec:safe-cap}.  In the case where only the Pre-Money SAFE converts, 
the payout to investor $1$ is 
\begin{align*}
 \payout_1(\{1\}) & =  \frac{\principal_1\scommon/\Cap_1}{\scommon + \principal_1\scommon/\Cap_1}(V- \principal_2)\\[10pt]
& = \frac{\principal_1}{\Cap_1 + \principal_1}(V- \principal_2)~. 
\end{align*}

The calculations for the Post-Money SAFE follow the description of Section~\ref{sec:post-money-cap}.
The number of shares issued is 
$$\sconverted{\Converted}_2 = (\principal_2/\Cap_2) ( \scommon + \sum_{i \in \Converted} \sconverted{\Converted}_i~)$$
that is, $\principal_2/\Cap_2$ of the total issuance after conversion. 

In the case $\Converted = \{2\}$, where only the 
Post-Money SAFE converts, 
the payout to player $2$ is the same fraction of the remaining value, that is, 
$\payout_2(\{1\}) =  (\principal_2/\Cap_2) (V - \principal_1)$. 

In the case where both SAFEs convert, we get 
that the payout is, similarly, $\payout_2(\emptyset) =  (\principal_2/\Cap_2) V$. 
The share issuance in this case is expressed in the equation 
\begin{align*} 
\sconverted{\{1,2\}}_2 & = (\principal_2/\Cap_2) ( \scommon +\sconverted{\{1,2\}}_1+ \sconverted{\{1,2\}}_2 )\\
& = (\principal_2/\Cap_2) ( \scommon + (\principal_1\scommon/\Cap_1)+ \sconverted{\{1,2\}}_2 )~.
\end{align*}
Calculating the actual share issuance to 
investor $2$ by solving the equation, we have 
$$\sconverted{\{1,2\}}_2 = \frac{\principal_2}{\Cap_2- \principal_2} \cdot \frac{\principal_1 + \Cap_1}{\Cap_1}  \cdot \scommon~.$$ 
It follows that the payout to the Pre-Money SAFE in this case is 
\begin{align*} 
\payout_1(\emptyset) & = \frac{\principal_1\scommon/\Cap_1}{ \scommon + (\principal_1\scommon/\Cap_1) + 
\frac{\principal_2}{\Cap_2- \principal_2} \cdot \frac{\principal_1 + \Cap_1}{\Cap_1}  \cdot \scommon}\cdot V \\
& = \frac{\principal_1}{\principal_1 + \Cap_1}\cdot \frac{\Cap_2  - \principal_2}{\Cap_2} \cdot V
\end{align*} 

We remark that the payouts just calculated do not fit within the Liquidity Event game model of Section~\ref{sec:game-model}. 

\begin{figure} 
\begin{center} 
\begin{tikzpicture}

\matrix[matrix of math nodes,every odd row/.style={align=right},every even row/.style={align=left},every node/.style={text width=1.5cm},row sep=0.2cm,column sep=0.2cm] (m) {
2&  1.77 \\ 

2&2\\

2& 2.34  \\ 

2.07 & 1.95 \\ 
};
\draw (m.north east) rectangle (m.south west);
\draw (m.north) -- (m.south);
\draw (m.east) -- (m.west);

\coordinate (a) at ($(m.north west)!0.25!(m.north east)$);
\coordinate (b) at ($(m.north west)!0.75!(m.north east)$);
\node[above=5pt of a,anchor=base] {Cash};
\node[above=5pt of b,anchor=base] {Convert};

\coordinate (c) at ($(m.north west)!0.25!(m.south west)$);
\coordinate (d) at ($(m.north west)!0.75!(m.south west)$);
\node[left=2pt of c,text width=1cm]  {Cash};
\node[left=2pt of d,text width=1cm]  {Convert};

\node[above=18pt of m.north] (firm b) {Investor 2};
\node[left=1.6cm of m.west,rotate=90,align=center,anchor=center] {Investor 1};

\end{tikzpicture}
\end{center} 
\caption{A liquidity game  mixing Pre-Money and Post-Money SAFEs\label{fig:mixed}} 
\end{figure}

Consider now the specific instance with, for the Pre-Money SAFE, $\principal_1 = 2$, $ \Cap_1 = 7$, 
for the Post-Money SAFE, $\principal_2 = 2$, $\Cap_2 = 4$, and value to be distributed $V = 8.2$. 
Then the above equations imply that we get the Liquidity game given in Figure~\ref{fig:mixed}. 
(The numbers are approximated to two decimal points.) 
This has a cycle of best responses 2:Cash $\rightarrow$ 1:Convert $\rightarrow$ 2:Convert $\rightarrow$ 1:Cash $\rightarrow$ 2:Cash. 
Thus,  this game has no pure strategy Nash equilibrium. 

This means that the combination of Pre-Money and Post-Money SAFEs can potentially create inherent conflicts amongst SAFE investors 
at the time of a Liquidity Event.  Interestingly, Y Combinator's Post-Money SAFE User Guide \cite{YC}  recommends against combining 
Post-Money and Pre-Money SAFEs, or combining SAFEs and convertible notes, the reason given being that this would 
require a more complex cap-table analysis. The analysis of the present section suggests that, some extra complexity aside, 
there is not an inherent problem in computing conversion amounts with such combinations. However, mixing of SAFE types may raise difficulties at the time of a Liquidity Event, that could require extra-contractual resolution.

\section{Related Work and Open Problems} \label{sec:concl} 

We have considered only pure strategy Nash equilibria in this work, motivated from the perspective 
that well designed convertible instruments should guarantee the existence of an optimum pure strategy equilibrium in order
to prevent inherent conflicts between different investors that might require extra-contractual resolution. 
From this point of view it is a positive result that pure strategy optima 
are guaranteed to exist in both Pre-Money and Post-Money SAFEs. However, mixed strategy equilibria 
may still be worth studying from a theoretical point of view.  We have required two different arguments 
for the two special cases where we have shown that pure strategy equilibria exist. Example~\ref{ex:noeq} shows that 
not all instances of our game model have a pure strategy equilibrium. 
An analysis of mixed strategy Nash equilibria, which always exist, might therefore help to give a unifying treatment. 

There exists literature on the computational complexity of computing Nash equilibria \cite{AGT}, 
generally focussed on computing mixed Nash equilibria, a problem that is known to be computationally
complex in general. Our polynomial time complexity results do not fit directly into this literature in that 
we do not represent the game in the input in matrix form: to do so for a Liquidity Event 
game with $n$ investors would require 
a representation already of size $2^n$. The fact that a problem has high complexity in general 
also does not inhibit the existence of 
low complexity instances. Our results show that the particular games we consider have
lower complexity than the general problem. 

Our analysis in this paper has assumed that the letter of the contracts will be applied in 
distributing funds to the investors. 
Conceivably, such a distribution according to contract may still meet with objections by some party. 
Depending on the jurisdiction, adjudication of such claims may result in court orders to modify the distribution,
on the grounds that some principle of fairness has legal priority over the contractual terms. 
The possibility of court intervention has not been considered in the present work, but 
may, for some jurisdictions, be worth considering. The area of bankruptcy theory and ``claims problems" 
\cite{THOMSON} provides many different approaches to the ``fair" division of competing claims
whose total exceeds the amount to be distributed. 

Another matter beyond the scope of our work is the question of whether  the Liquidity Event should proceed at all.  
SAFE contracts distinguish between Liquidity Events and Dissolution Events (in which the company is liquidated), 
and state a simpler distribution rule for Dissolution Events (a simple return of principal), 
but the Liquidity Event rules could sensibly be used for  Dissolution Events also. 
There is literature \cite{Schwartz93} on game theoretic aspects of negotiations between managers of a 
company facing liquidation and its creditors. The choices in this case concern 
offers and acceptances of terms for restructuring the obligations of the company in order 
to maintain it as a going concern, versus liquidation of the company. 
Game theoretic reasoning has also been applied in structuring proposals to creditors 
concerning the legal jurisdiction under which a restructuring is to be conducted \cite{Turner}.  
These works are orthogonal to our concerns in this 
paper, where it has already been determined that the Liquidity Event game is to be played.  

There has also been game theoretic analysis of convertible notes \cite{BS80} in which the game is 
played between the issuing company and an investor, rather than between different investors, as in our analysis. 
In this work, both the company and the investor have options: the company may choose to convert the note to shares once the share price reaches a certain level, 
and the investor has an option, after the maturity date, to either recoup principal or convert to shares. 
The analysis of the game concentrates on the timing of the decision to convert, and is used as a basis for pricing the convertible note. The 
conversion price for the notes is generally an amount fixed in advance, rather than an amount that depends on the 
share price of the company, as in SAFE notes. SAFEs also  differ in that the underlying shares generally do not trade 
on public markets, and the condition allowing conversion is an equity round, liquidity event or dissolution, rather than 
a particular share price being reached.  

SAFE notes are equity-like instruments, in that they pay no interest and are intended to convert to equity, 
whereas convertible bonds are more debt-like, in that they pay interest but have an option to convert. 
However, potentially a similar analysis of the conversion decision in liquidity events may apply to convertible bonds, 
in which we would interpret $\principal_i$ as the principal plus remaining interest due, rather than as simply the principal. The details of specific instruments used in practice would need to be investigated to validate this intuition, however.

\bibliographystyle{alpha}
\bibliography{refs}

\end{document}